\documentclass[aps,prd, twocolumn]{revtex4}
\usepackage{amsmath,amssymb,amsfonts}
\usepackage{graphicx}
\usepackage{enumerate} 
\usepackage{colordvi} 
\usepackage{bm}

\newcommand{\bfv}{\mbox{\boldmath$v$}}

\newcommand{\bfr}{\mbox{\boldmath$r$}}

\newcommand{\sigmav}{\sigma_{\rm v}}
\newcommand{\Pdd}{P_{\delta\delta}}
\newcommand{\Pdv}{P_{\delta \theta}}
\newcommand{\Pvv}{P_{\theta\theta}}

\begin{document}
\title{Forecasting the Cosmological Constraints with Anisotropic 
Baryon Acoustic Oscillations from Multipole Expansion}
\vfill
\author{Atsushi Taruya$^{1,2}$, Shun Saito$^{3,4}$, Takahiro Nishimichi$^2$}

\bigskip
\address{$^1$Research Center for the Early Universe, School of Science, 
The University of Tokyo, Bunkyo-ku, Tokyo 113-0033, Japan}
\address{$^2$Institute for the Physics and Mathematics of the Universe, 
The University of Tokyo, Kashiwa, Chiba 277-8568, Japan}
\address{$^3$Department of Physics, The University of Tokyo, 
113-0033, Japan}
\address{$^4$Department of Astronomy, University of California Berkley, 
California 94720, USA}
\bigskip
\date{\today}
%
\begin{abstract}
Baryon acoustic oscillations (BAOs) imprinted in the galaxy power spectrum 
can be used as a standard ruler to determine angular diameter distance 
and Hubble parameter at high redshift galaxies. Combining redshift 
distortion effect which apparently distorts the galaxy clustering pattern, 
we can also constrain the growth rate of large-scale structure formation. 
Usually, future forecast for constraining these parameters from galaxy 
redshift surveys has been made with a full 2D power 
spectrum characterized as function of wavenumber $k$ and directional 
cosine $\mu$ between line-of-sight direction and wave vector, i.e., 
$P(k,\mu)$. Here, we apply the multipole expansion to the full 2D 
power spectrum, and discuss how much cosmological information can be  
extracted from the lower-multipole spectra, taking a proper account of the 
non-linear effects on gravitational clustering and redshift distortion. 
The Fisher matrix analysis reveals that compared to the analysis with full 
2D spectrum, a partial information from the monopole and 
quadrupole spectra generally degrades the constraints by a factor of 
$\sim1.3$ for each parameter. The additional information from the 
hexadecapole spectrum helps to improve the constraints, which lead to 
an almost comparable result expected from the full 2D spectrum. 
\end{abstract}

\maketitle

\maketitle
\section{Introduction}

Baryon acoustic oscillations (BAOs) imprinted on the clustering of 
galaxies are now recognized as a powerful cosmological probe to 
trace the expansion history of the Universe 
\cite{Eisenstein:2005su,Percival:2007yw,Percival:2009xn}. 
In particular, the BAO measurement via a spectroscopic survey can 
provide a way to 
simultaneously determine the angular diameter distance $D_A$ 
and Hubble parameter $H$ at given redshift of galaxies through the  
cosmological distortion, known as Alcock-Paczynski effect 
(e.g., \cite{Alcock_Paczynski:1979,Seo:2003pu,Blake:2003rh,Shoji:2008xn,
Padmanabhan:2008ag}). 
Further, measuring the clustering anisotropies caused by the redshift 
distortion due to the peculiar velocity of galaxies, we can also probe 
the growth history of structure formation 
(e.g., \cite{Linder:2007nu,Guzzo:2008ac,Yamamoto:2008gr,Song:2008qt}), 
characterized by the growth-rate parameter $f\equiv d\ln D/d\ln a$, with 
quantities $D$ and $a$ being linear growth factor and the scale factor of 
the Universe, respectively.

With the increased number of galaxies and large survey volumes,  
on-going and future spectroscopic galaxy surveys such as 
Baryon Oscillation Spectroscopic 
Survey (BOSS) \cite{Schlegel:2009hj}, 
Hobby-Eberly Dark Energy Experiment (HETDEX) \cite{Hill:2008mv}, 
Subaru Measurement of Imaging and Redshift equipped with 
Prime Focus Spectrograph (SuMIRe-PFS), and EUCLID/JDEM 
\cite{Beaulieu:2010qi,Gehrels:2010fn} 
aim at precisely measuring 
the acoustic scale of 
BAOs as a standard ruler. These surveys will cover the wide redshift ranges, 
$0.3\lesssim z\lesssim 3.5$, and provide a precision data of 
the redshift-space power spectrum with an accuracy of a percent level over 
the scales of BAOs.

In promoting these gigantic surveys, 
a crucial task is a quantitative forecast for the size 
of the statistical errors on the parameters 
$D_A$, $H$ and $f$ in order to clarify the scientific benefits 
as well as to explore the optimal survey design. 
The Fisher matrix formalism is a powerful 
tool to investigate these issues, and it enables us to quantify the 
precision and the correlation between multiple parameters 
(\cite{Seo:2003pu,Seo:2007ns,White:2008jy,Shoji:2008xn},  
especially for measuring $D_A$, $H$ and $f$). 
So far, most of the works on the parameter forecast study 
have focused on the potential power of the BAO measurements, and 
attempt to clarify the achievable level of the precision for the 
parameter estimation. For this purpose, they sometimes assumed a rather 
optimistic situation that a full shape of the redshift-space 
power spectrum, including the clustering anisotropies due to the 
redshift distortion, is available in both observation and theory.

In this paper, we are particularly concerned with the parameter 
estimation using a partial information of the anisotropic BAOs 
from a practical point-of-view. 
In redshift space, the power spectrum obtained from the spectroscopic 
measurement is generally described in the two dimension, and is 
characterized as functions of $k$ and $\mu$, where $k$ is the 
wavenumber and $\mu$ 
is the directional cosine between the line-of-sight direction and $k$ 
\footnote{Throughout the paper, we work with the distant-observer 
approximation, and neglect the angular dependence of the 
line-of-sight direction, relevant for the high-redshift galaxy surveys.}. 
While most of the forecast study is concerned with a full 2D power 
spectrum, the multipole expansion of redshift-space power spectrum 
has been frequently used in the data analysis 
to quantify the clustering anisotropies.  
Denoting the power spectrum by $P(k,\mu)$, we have
\begin{align}
&P(k,\mu)=\sum_{\ell=0}^{\rm even} P_\ell(k)\,\mathcal{P}_\ell(\mu) 
\label{eq:multipole_pk}
\end{align}
with the function $\mathcal{P}_\ell$ being the Legendre polynomials. 
Although the analysis with full 2D spectrum will definitely play an 
important role as improving the statistical signal, most of the 
recent cosmological data 
analysis has focused on the angle-averaged power spectrum $(\ell=0)$, 
i.e., monopole spectrum, and a rigorous analysis with full 2D spectrum is 
still heavy task due to the time-consuming covariance estimation 
(e.g., \cite{Okumura:2007br,Takahashi:2009bq,Cabre:2008sz}).

In linear theory, the redshift-space power spectrum is simply written as 
$P(k,\mu)=(1+\beta\,\mu^2)^2P_{\rm gal}(k)$, where $\beta=f/b$ with $b$ being 
the linear bias parameter, and $P_{\rm gal}$ 
is the galaxy power spectrum in real space 
\cite{Kaiser:1987qv,1992ApJ385L5H,Hamilton:1997zq}. 
Then, the non-vanishing components arises only from the monopole ($\ell=0$), 
quadrupole ($\ell=2$) and hexadecapole spectra ($\ell=4$). 
That is, cosmological information contained in the $\ell=0$, 
$2$ and $4$ moments is equivalent to the whole information 
in the full 2D power spectrum. Observationally, however, this is only 
the case when we a priori know the cosmological distance to the galaxies. 
The Alcock-Paczynski effect can induce non-trivial clustering anisotropies,  
which cannot be fully characterized by the lower multipole spectra, 
in general.  Further, in reality, linear theory description cannot be 
adequate over the scale of the BAOs, and the non-linear effects of 
the redshift distortion as well as the gravitational clustering 
must be accounted for a proper comparison with observation. 
These facts imply that non-vanishing multipole spectra higher 
than $\ell>4$ generically appear, and a part of the cosmological 
information might be leaked into those higher multipole moments. 
An important question is how much amount of the 
cosmological information can be robustly extracted from the 
lower multipole spectra instead of the full 2D spectrum. 
In the light of this, Ref.~\cite{Padmanabhan:2008ag} recently 
examined a non-parametric method to constrain $D_A$ and $H$ from the 
monopole and quadrupole spectra, and numerically estimate the size of 
errors (see also Ref.~\cite{TocchiniValentini:2011mt} for the 
estimation of growth-rate parameter).

Here, as a complementary and comprehensive approach, 
we will investigate this issue 
based on the Fisher matrix formalism, and derive the useful 
formulae for parameter forecast using the multipole power spectra. 
We then explore the potential power of the lower multipole 
spectra on the cosmological constraints, 
particularly focusing on the parameters $D_A$, $H$ and $f$. 
To do so,  we consider the Figure-of-Merit (FoM) and 
Figure-of-Bias (FoB) for these parameters, and investigate their
dependence on the assumptions for the number density of galaxies, 
the amplitude of clustering bias, the maximum wavenumber used for 
the parameter estimation.

In Sec.~\ref{sec:Fisher_formalism}, we present the Fisher matrix 
formalism for cosmological parameter estimation from the multipole 
power spectra. Sec.~\ref{sec:model_assumption} deals with the model 
of redshift-space power spectrum and the assumptions used in the Fisher 
matrix analysis. Then, in Sec.~\ref{sec:results}, 
the results for FoM and FoB are shown, and the sensitivity of the results 
to the assumptions and choice of the parameters is discussed in 
greater details. Finally, Sec.~\ref{sec:summary} briefly summarize our 
present work. 

Throughout the paper, we assume a flat Lambda cold dark matter (CDM) model, 
and the fiducial model parameters are chosen based on the five-year 
WMAP results \cite{Komatsu:2008hk}: 
$\Omega_{\rm m}=0.279$, 
$\Omega_{\Lambda}=0.721$, $\Omega_{\rm b}=0.0461$, $h=0.701$,
$n_s=0.96$, $A_s=2.19\times10^{-9}$. 

\section{Fisher matrix formalism}
\label{sec:Fisher_formalism}

In this section, we present the basic formulae for Fisher matrix 
analysis in estimating the statistical error and systematic biases for 
cosmological parameters from the multipole power spectra.

Let us first derive the expression for Fisher matrix 
relevant for the power 
spectrum analysis. The definition of the Fisher matrix is given by 
\begin{align}
F_{ij}=-\left\langle\frac{\partial^2
\ln \mathcal{L}}{\partial\theta_i\partial \theta_j}\right\rangle,
\label{eq:def_Fij}
\end{align}
where $\theta_i$ denotes the parameter, and the quantity 
$\mathcal{L}$ is the likelihood function. For the parameter 
estimation study with the multipole spectrum, $P_\ell(k)$, 
the likelihood function is usually taken in the form as 
\begin{align}
\mathcal{L}\propto \exp\left[-\frac{1}{2}\sum_{m,n}\sum_{\ell,\ell'}
\Delta P_\ell(k_m)\left[C^{\ell\ell'}(k_m,k_n)\right]^{-1}\Delta P_{\ell'}(k_n)
\right],
\label{eq:likelihood}
\end{align}
where we define
\begin{eqnarray}
\Delta P_\ell(k)&\equiv& \widehat{P}_\ell(k) - P_\ell(k),
\nonumber\\
C^{\ell\ell'}(k_m,k_n)&\equiv&\langle\Delta P_\ell(k_m)\,
\Delta P_{\ell'}(k_n)\rangle. 
\nonumber
\end{eqnarray}
The quantities $\widehat{P}_\ell(k)$ and $P_\ell(k)$ 
respectively denote the observed estimate and theoretical template 
for the multipole power spectrum.

Substituting Eq.~(\ref{eq:likelihood}) into the definition 
(\ref{eq:def_Fij}), the leading-order evaluation of the Fisher matrix 
leads to (e.g., \cite{Yamamoto:2002bc,Tegmark:1997rp}): 
\begin{equation}
F_{ij}\simeq \sum_n\sum_{\ell,\ell'} 
\frac{\partial P_{\ell}(k_n)}{\partial \theta_i}\,
\left[\mbox{Cov}^{\ell\ell'}(k_n)\right]^{-1}\,
\frac{\partial P_{\ell}(k_n)}{\partial \theta_j},
\end{equation}
where we have assumed that the covariance is approximately characterized 
by the Gaussian statistic, and is written as 
$C^{\ell\ell'}(k_m,k_n) = \mbox{Cov}^{\,\ell\ell'}(k_n)\,\delta_{mn}$.

Adopting the power spectrum estimation by Ref.~\cite{Feldman:1993ky}, 
the analytic expression for the quantity $\mbox{Cov}^{\ell\ell'}(k_n)$ 
can be found in Ref.~\cite{Yamamoto:2005dz} [see Eq.~(25) of their paper]: 
\begin{align}
&\mbox{Cov}^{\ell\ell'}(k_n)=\frac{2}{V_n}\,\frac{(2\ell+1)(2\ell'+1)}{2}\,
\nonumber\\
&\qquad\times\int_{-1}^1 d\mu \frac{\mathcal{P}_\ell(\mu)\mathcal{P}_{\ell'}(\mu)}{
\int d^3\bfr \,\,\overline{n}(\bfr)^2
[1+\overline{n}(\bfr)\,P^{\rm(S)}(k_n,\mu)\,]^{-2}}
\label{eq:cov}
\end{align}
with $\mathcal{P}_\ell(\mu)$ being the Legendre polynomial\footnote{ 
Here, we use the standard 
notation for the multipole expansion of redshift power spectra given by  
(\ref{eq:multipole_pk}), which differs from the definition of
Ref.~\cite{Yamamoto:2005dz} }.  
The quantity $V_n$ is the volume element of the thin shell 
in the Fourier space, i.e., $V_n=4\pi^2k_n^2 dk_n/(2\pi)^3$, 
which corresponds to $\Delta V_k/(2\pi)^3$ in the notation of 
Ref.~\cite{Yamamoto:2005dz}.

Now, to simplify the formula, we consider the homogeneous 
galaxy samples, which implies $\overline{n}(\bfr)=\overline{n}=\mbox{const}$. 
In this case, the denominator in the integrand of Eq.~(\ref{eq:cov}) is 
simplified as 
\begin{align}
&\int d^3\bfr \,\,\overline{n}(\bfr)^2
[1+\overline{n}(\bfr)\,P(k,\mu)\,]^{-2}
\nonumber\\
&\qquad\qquad\qquad\qquad=
V_s\,\,\left\{P(k,\mu)+\frac{1}{\overline{n}}\right\}^{-2}, 
\end{align}
where $V_s$ denotes the 
survey volume. Then, taking the continuum limit, 
the expression for Fisher matrix can be recast as 
\begin{align}
&F_{ij}= \frac{V_s}{4\pi^2}\int_{k_{\rm min}}^{k_{\rm max}} 
dk\,k^2\,\,\sum_{\ell,\ell'} 
\frac{\partial P_{\ell}(k)}{\partial \theta_i}
\left[\widetilde{\mbox{Cov}}^{\ell\ell'}(k)\right]^{-1}
\frac{\partial P_{\ell}(k)}{\partial \theta_j},
\label{eq:formula_F_ij}
\end{align}
with the reduced covariance matrix $\widetilde{\mbox{Cov}}^{\ell\ell'}(k)$ 
given by
\begin{align}
&\widetilde{\mbox{Cov}}^{\ell\ell'}(k)=\frac{(2\ell+1)(2\ell'+1)}{2}\,
\nonumber\\
&\qquad
\times\int_{-1}^1d\mu\,\mathcal{P}_\ell(\mu)\,\mathcal{P}_{\ell'}(\mu)\,
\left[P(k,\,\mu)+\frac{1}{\overline{n}}\right]^2.
\label{eq:cov_formula}
\end{align}
Here, the range of integration $[k_{\rm min}, k_{\rm max}]$ should be 
chosen through the survey properties and/or limitation of theoretical 
template, and, in particular, the minimum wave number is limited to 
$2\pi/V_s^{1/3}$.

Eq.~(\ref{eq:formula_F_ij}) with (\ref{eq:cov_formula}) is 
the formula for the Fisher matrix used in the parameter estimation 
with multipole power spectra. This can be compared with the standard 
formula for full 2D power spectrum 
(e.g., \cite{Tegmark:1997rp,Seo:2003pu,Shoji:2008xn}):
\begin{align}
&F_{ij}^{\rm (2D)}= \frac{V_s}{4\pi^2}\int_{k_{\rm min}}^{k_{\rm max}} 
dk k^2\int_{-1}^1d\mu\,
\frac{\partial P(k,\mu)}{\partial \theta_i}
\left\{P(k,\mu)+\frac{1}{\overline{n}}\right\}^{-2}
\nonumber\\
&\qquad\qquad\times\frac{\partial P(k,\mu)}{\partial \theta_j}
\label{eq:2D_formula_F_ij}
\end{align}
That is, the full 2D information obtained through the integral over 
directional cosine $\mu$ in Eq.(\ref{eq:2D_formula_F_ij}) is 
replaced with the summation over all multipoles in the 
new formula (\ref{eq:formula_F_ij}). Thus, truncating the summation at 
a lower multipole generally leads to the reduction of the amplitude in 
Fisher matrix, and as a result,  the statistical errors of the parameter 
$\theta_i$ marginalized over other parameters, given by 
$\Delta\theta_i=\sqrt{\{F^{-1}\}_{ii}}$, is expected to become larger.

The Fisher matrix formalism also provides a simple way to estimate 
the biases in the best-fit parameters caused by the 
incorrect template for the multipole power spectra $P_{\ell}^{\rm wrong}(k)$. 
To derive the formula for systematic bias, we replace the template 
power spectrum $P_\ell(k)$ in the likelihood function (\ref{eq:likelihood}) 
with the incorrect one 
$P_{\ell}^{\rm wrong}(k)$. We denote this likelihood function by 
$\mathcal{L}'$. Assuming that the size of the biases are basically 
small, the (biased) best-fit values can be estimated from the extremum 
of the Likelihood function $\mathcal{L}'$ 
by expanding the expression of the extremum around the fiducial parameters:
\begin{align}
0=\frac{\partial\ln\mathcal{L'}}{\partial\theta_j}\simeq
\left.\frac{\partial\ln\mathcal{L'}}{\partial\theta_j}\right|_{\rm fid}+
\left.\sum_i \frac{\partial \ln\mathcal{L'}}{\partial\theta_i\partial\theta_j}
\right|_{\rm fid}\delta \theta_i,
\label{eq:extremum_likelihood}
\end{align}
where the quantities with subscript $_{\rm fid}$ stand for the one 
evaluated at the fiducial parameters, and the $\delta\theta_i$ means the 
deviation of the best-fit value from the fiducial parameter. 
Then, taking the ensemble average of the above expressions and using 
the definition of the Fisher matrix, we obtain
\begin{align}
\delta \theta_i = -\sum_j(F')^{-1}_{ij}s_j,
\label{eq:systematic_bias}
\end{align}
where the Fisher matrix $F'_{ij}$ is the same one as given by 
Eq.~(\ref{eq:formula_F_ij}), but is evaluated using incorrect 
power spectra $P_{\ell}^{\rm wrong}(k)$. The vector $s_j$ is 
 \begin{align}
 s_j = \frac{V_s}{4\pi^2}\int_{k_{\rm min}}^{k_{\rm max}} 
dk \,k^2 \sum_{\ell,\ell'} P_{\ell}^{\rm sys}(k)
 \left[\widetilde{\rm Cov}^{\ell\ell'}(k)\right]^{-1}\,
 \frac{\partial P_{\ell'}^{\rm wrong}(k)}{\partial \theta_j}.
\label{eq:vector_s}
 \end{align}
Here, the multipole power spectrum $P_{\ell}^{\rm sys}(k)$ denotes the 
systematic difference between correct and incorrect model of 
multipole power spectra, 
$P_{\ell}^{\rm sys}(k)=P_{\ell}^{\rm wrong}(k)-P_{\ell}^{\rm true}(k)$. 
In deriving the above expression, we have used the fact that the extremum 
of the likelihood function vanishes only when the correct template for the 
multipole power spectrum is applied.

Notice that similar but essentially different formula for systematic biases 
is obtained in the cases using the full 2D power spectrum. 
It is formally expressed as Eq.~(\ref{eq:systematic_bias}), 
but the Fisher matrix $F'_{ij}$ 
is now replaced with Eq.~(\ref{eq:2D_formula_F_ij}) evaluated using 
the incorrect 2D spectrum $P^{\rm wrong}(k,\mu)$. 
Further, the vector $s_j$ should be replaced with the one for 
the full 2D spectrum (e.g., \cite{Saito:2009ah,Taruya:2010mx}):  
 \begin{align}
& s_j^{\rm (2D)} = 
\frac{V_s}{4\pi^2}\int_{k_{\rm min}}^{k_{\rm max}}  
dk \,k^2 \int_{-1}^1 d\mu \,\,P^{\rm sys}(k,\mu)
\nonumber\\
&\qquad\qquad\times
 \left[P_{\ell'}^{\rm wrong}(k,\mu)+\frac{1}{\overline{n}_{\rm g}}\right]^{-2}\,
\frac{\partial P^{\rm wrong}(k,\mu)}{\partial \theta_j}.
 \end{align}

Finally, all the formulae derived in this section 
ignore the non-Gaussian contributions to the likelihood and 
covariances, which would be sometimes important in practice.  
The extension of the formulae to include the non-Gaussian 
contributions is straightforward, and will be considered elsewhere. 
For the effects of non-Gaussian contributions to the parameter 
estimation study especially focusing on BAOs, 
several works have been recently done based on the numerical and 
analytical treatments 
\cite{Takahashi:2009bq,Takahashi:2009ty,Neyrinck:2007bp}.

\begin{figure}[t]
\begin{center}
\includegraphics[width=8.0cm,angle=0]{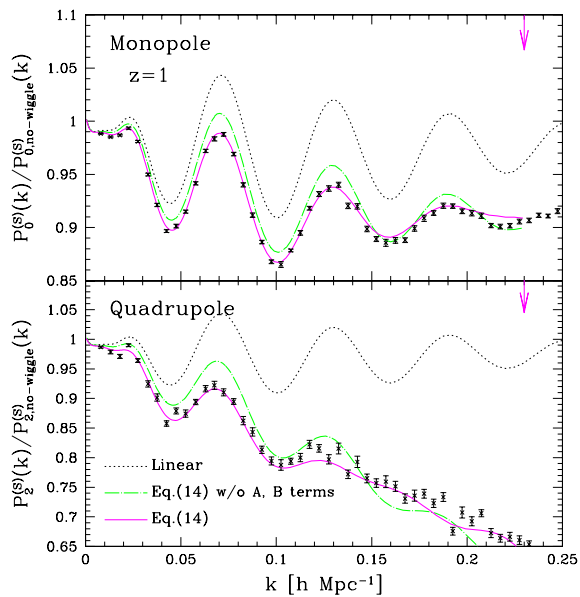}
\end{center}

\vspace*{-0.5cm}

\caption{Monopole (top) and quadrupole (bottom) moments of matter 
  power spectra in redshift space at $z=1$. The results are  
  divided by the smooth reference spectrum, 
  $P_{\ell,{\rm no\mbox{-}wiggle}}^{\rm (S)}$, and are compared with 
  the N-body results (symbols) 
  taken from the {\tt wmap5} simulations of Ref.~\cite{Taruya:2009ir}. 
  The reference spectrum $P_{\ell,{\rm no\mbox{-}wiggle}}^{\rm (S)}$ is 
  calculated from the no-wiggle approximation of the linear transfer 
  function \cite{Eisenstein:1997ik} with the linear theory of the Kaiser 
  effect taken into account. 
  Solid and dot-dashed lines represent the results of improved PT 
  calculations based on the model of redshift
  distortion (\ref{eq:new_model}), but the terms $A$ and $B$ are ignored 
  in the dot-dashed lines. In both cases, the one-dimensional velocity 
  dispersion $\sigmav$ was determined by fitting the predictions 
  to the N-body simulations, using the data below the wavenumber 
  indicated by the vertical arrow. The best-fit values of $\sigmav$ are 
  $\sigmav=395\,$km\,s$^{-1}$ and $285\,$km\,s$^{-1}$, with and without the 
  $A$ and $B$ terms, respectively. 
\label{fig:ratio_pkred}}
\end{figure}

\section{Model and Assumptions}
\label{sec:model_assumption}

Given the formulae for Fisher matrix analysis,  
we now move to the discussion on the parameter forecast study using the 
multipole power spectra, and compare the results with those obtained from 
the full 2D spectrum. Before doing this, in this section, we briefly 
describe the model and assumptions for redshift-space power spectrum 
relevant for spectroscopic measurement of BAOs.

In redshift space, clustering statistics 
generally suffer from the two competitive effects, i.e., enhancement 
and suppression of clustering amplitude, referred to as the 
Kaiser and  Finger-of-God effects, respectively. While the 
Kaiser effect comes from the coherent motion of the 
mass (or galaxy), the Finger-of-God effect 
is mainly attributed to the virialized random motion of the mass residing 
at a halo. On weakly non-linear regime,  
a tight correlation between 
velocity and density fields still remains, and a mixture of Kaiser and 
Finger-of-God effects is expected to be significant. Thus, a careful 
treatment is needed for accurately modeling anisotropic 
power spectrum.

Recently, 
we have presented an improved prescription for matter 
power spectrum in redshift space taking account of both the 
non-linear clustering and redshift distortion \cite{Taruya:2010mx}. 
Based on the perturbation theory calculation, 
the model can give an excellent 
agreement with results of N-body simulations, and  a percent level 
precision is almost achieved over the scales of our interest on BAOs. 
The full 2D power spectrum of this model is very similar to the 
one proposed by Ref.~\cite{Scoccimarro:2004tg}, but includes the corrections: 
\begin{align}
&P(k,\mu)=e^{-(k\mu\,f\sigmav)^2}\,\Bigl\{\Pdd(k)
+2\,f\,\mu^2\,\Pdv(k)
\nonumber\\
&\qquad+f^2\,\mu^4\,\Pvv(k)+A(k,\mu;f)+B(k,\mu;f)
\Bigr\}  
\label{eq:new_model}
\end{align}
with the quantity $f$ being the growth-rate parameter. 
Here, the power spectra $\Pdd$, $\Pvv$ and $\Pdv$ denote the 
auto power spectra of density and velocity divergence, and their
cross power spectrum, respectively. The velocity divergence 
$\theta$ is defined by $\theta\equiv-\nabla \bfv/(aHf)$. 
The quantity $\sigmav$ denotes the one-dimensional 
velocity dispersion\footnote{The definition of velocity dispersion $\sigmav$ 
adopted in this paper differs from the one commonly used in the literature by 
a factor of $f$, but coincides with those in 
Refs.~\cite{Scoccimarro:2004tg,Taruya:2010mx}.}, 
and the exponential prefactor characterizes 
the damping behavior by the Finger-of-God effect. For the purpose 
to model the shape and structure of BAOs in power spectrum, 
$\sigmav$ may be treated as a free parameter, and determine it by 
fitting the predictions to the observations.

A salient property of the model (\ref{eq:new_model}) is the presence of 
the terms $A$ and $B$, which represent the higher-order couplings between 
velocity and density fields, usually neglected in the 
phenomenological models of redshift distortion. The explicit expressions 
for these terms are derived based on the standard treatment of perturbation 
theory, and the results are presented in Ref.~\cite{Taruya:2010mx}. A detailed 
investigation in our previous paper \cite{Taruya:2010mx} reveals that 
the corrections $A$ and $B$ can give an important contribution to 
the acoustic structure of BAOs over the scales 
$k\sim0.2h$Mpc$^{-1}$, which give rise to a slight uplift in the amplitude 
of monopole and quadrupole spectra. With the improved treatment of 
the perturbation theory to compute $\Pdd$, $\Pvv$ and $\Pdv$ 
(e.g., \cite{Crocce:2007dt,Taruya:2009ir}), 
the model (\ref{eq:new_model}) can give a better prediction than 
the existing models of redshift distortion. 
Fig.~\ref{fig:ratio_pkred} plots the illustrated example showing that 
the model (\ref{eq:new_model}) reproduces the N-body results of 
monopole and quadrupole spectra quite well, and the precision of the 
agreement between prediction and simulation reaches a percent-level. 
Hence, in this paper, we adopt the model (\ref{eq:new_model}) as a 
fiducial model for matter power spectrum in redshift space.

Note that the model (\ref{eq:new_model}) 
generically produces the non-vanishing higher multipole spectra 
of $\ell>4$, due to the damping factor, $e^{-(k\mu\,f\sigmav)^2}$. 
Furthermore, the corrections $A$ and $B$ are expanded as 
power series of $\mu$, which include the powers up to $\mu^6$ for 
the $A$ term, $\mu^8$ for the $B$ term. This indicates that the 
corrections additionally contribute to the higher multipoles, at least, 
up to $\ell=8$. In this sense, the model (\ref{eq:new_model}) provides 
an interesting testing ground to estimate the extent to which 
the useful cosmological information can be obtained from 
the lower-multipole spectra.

Then, assuming the linear galaxy bias in real space, 
$\delta_{\rm gal}=b\delta_{\rm mass}$, the redshift-space power 
spectrum for galaxies becomes 
\begin{align}
&P_{\rm gal}(k,\mu)=e^{-(k\mu\,f\sigmav)^2}\,b^2\,\Bigl\{\Pdd(k)
+2\,\beta\,\mu^2\,\Pdv(k)
\nonumber\\
&\qquad+\beta^2\,\mu^4\,\Pvv(k)+b\,A(k,\mu;\beta)+b^2\,B(k,\mu;\beta)
\Bigr\}   
\label{eq:new_pk_gal}
\end{align}
with $\beta=f/b$. The linear deterministic bias may be too simplistic
assumption, 
and the effects of non-linearity and 
stochasticity in the galaxy bias might be non-negligible 
\cite{Okumura:2010sv,Jeong:2008rj,Saito:2010pw}. 
Our primary concern here is the qualitative aspects of the parameter 
estimation using the multipole spectra, based on a physically plausible 
model of redshift distortion. Since the galaxy bias itself does not 
produce additional clustering anisotropies, we simply adopt the linear bias 
relation for illustrative purpose.

Finally, notice that in addition to the clustering anisotropies caused by 
the peculiar velocity of galaxies, the observed galaxy power spectrum 
defined in comoving space further exhibits anisotropies induced by the 
Alcock-Paczynski effect.  This is modeled as
\begin{equation}
P_{\rm obs}(k,\mu) = \frac{H(z)}{H_{\rm fid}(z)}
\left\{ \frac{D_{A,{\rm fid}}(z)}{D_A(z)}\right\}^2\,\,
P_{\rm gal}(q,\nu), 
\label{eq:pk_model} 
\end{equation}
where the quantity $P_{\rm gal}(q,\nu)$ at the right-hand-side represents 
the template for the redshift-space power spectrum in the absence of 
cosmological distortion,  i.e., Eq.(\ref{eq:new_pk_gal}). 
The comoving wavenumber $k$ and the directional cosine $\mu$ measured 
with the underlying cosmological model are related to the true ones 
$q$ and $\nu$ by the Alcock-Paczynski effect through 
(e.g., \cite{Ballinger:1996cd,Magira:1999bn,Padmanabhan:2008ag})
\begin{align}
&q=k\,\left[
\left(\frac{D_{A,{\rm fid}}}{D_A}\right)^2+
\left\{\left(\frac{H}{H_{\rm fid}}\right)-
\left(\frac{D_{A,{\rm fid}}}{D_A}\right)^2\right\}\mu^2
\right]^{1/2},
\\
&\nu= \left(\frac{H}{H_{\rm fid}}\right)\,\mu
\nonumber\\
&\qquad\,\times\left[
\left(\frac{D_{A,{\rm fid}}}{D_A}\right)^2+
\left\{\left(\frac{H}{H_{\rm fid}}\right)-
\left(\frac{D_{A,{\rm fid}}}{D_A}\right)^2\right\}\mu^2
\right]^{-1/2}, 
\end{align}
The quantities $D_{A,{\rm fid}}$ and $H_{\rm fid}$ are the fiducial 
values of the angular diameter distance and Hubble parameter at 
a given redshift slice.

\section{Results}
\label{sec:results}

In what follows, for illustrative purpose, we consider the 
hypothetical galaxy survey of the volume $V_s=4h^{-3}$Gpc$^3$ at $z=1$, 
and examine how well we can constrain the distance information and 
growth-rate parameter, $D_A$, $H$, and $f$, from the low-multipole 
power spectra. We set the number density of galaxies,  
linear bias parameter and velocity dispersion to 
$\overline{n}=5\times10^{-4}h^3$Mpc$^{-3}$,  
$b=2$ and $\sigmav=395$km s$^{-1}$.  These values are used in 
the Fisher analysis as a canonical setup, but we also examine the variants 
of these parameter set to study the sensitivity of the 
forecast results. Note that the depth and the volume of the
survey considered here roughly match those of a stage III class 
survey defined by the Dark Energy Task Force (DETF) \cite{Albrecht:2006um}. 

To compute the Fisher matrix adopting the model of redshift-space power 
spectrum, Eq.~(\ref{eq:new_pk_gal}), we just follow the procedure 
in Ref.~\cite{Taruya:2010mx} to calculate the redshift-space power spectra.  
That is, we use the improved PT developed 
by Ref.~\cite{Taruya:2009ir,Taruya:2007xy} 
to account for a dominant contribution of the non-linear gravity 
to the power spectra $\Pdd$, $\Pdv$ and $\Pvv$, and to adopt 
standard PT for small but non-negligible corrections of 
$A$ and $B$ terms. Detailed comparison with 
N-body simulations \cite{Taruya:2009ir,Taruya:2010mx} showed that 
this treatment can work well, and in our fiducial set of 
cosmological parameters, the model can give a percent-level precision 
at least up to the wavenumber $k\leq0.2h$Mpc$^{-1}$ at $z=1$.

Number of free parameters in the subsequent Fisher analysis is five in 
total, i.e.,  $D_A$, $H$, and $f$, in addition to the parameters 
$b$ and $\sigmav$. Other cosmological parameters such as 
$\Omega_{\rm m}$ or $\Omega_{\rm b}$ are kept fixed. 
We assume that the cosmological model dependence of the power 
spectrum shape is perfectly known a priori from the precision CMB 
measurement by PLANCK \cite{Planck}. The influence of the uncertainty in 
the power spectrum shape is discussed in Sec.~\ref{subsec:bias_prior} 
in detail.

\subsection{Two-dimensional errors}

\begin{figure}[t]
\begin{center}

\vspace*{-0.5cm}

\hspace*{-0.75cm}
\includegraphics[width=9.8cm,angle=0]{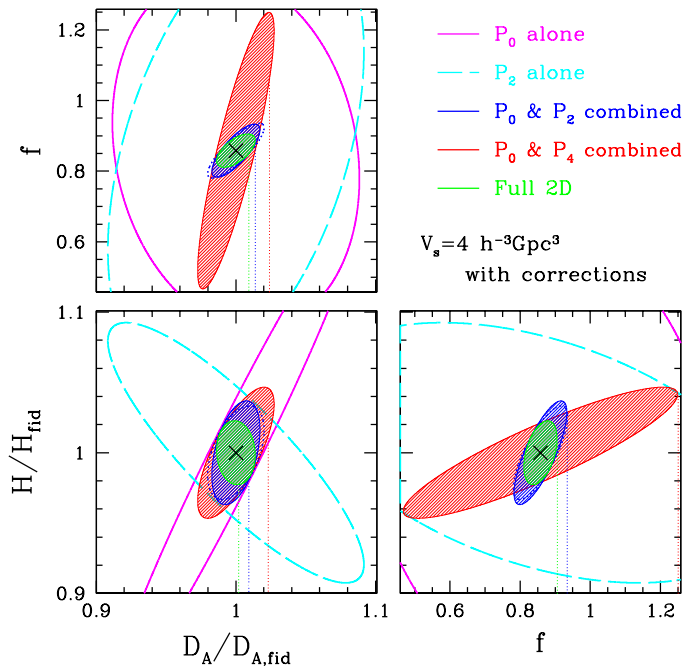}
\end{center}

\vspace*{-0.8cm}

\caption{Two dimensional contours of $1$-$\sigma$ ($68\%$CL) errors 
on $(D_A,H)$ (bottom-left), $(D_A,f)$ (top-left), and $(f,H)$ (bottom-right), 
assuming a stage-III class survey with $V_s=4h^{-3}$Gpc$^3$ at $z=1$.  
In each panel, magenta solid and cyan dashed lines respectively 
indicate the forecast constraints coming from the monopole 
($P_0$) and quadrupole $(P_2)$ spectrum alone, while the blue and red 
shaded region represent the combined constraints from $P_0$ and $P_2$, and 
$P_0$ and $P_4$, respectively. 
The inner green shaded region is the results coming from the full 2D 
spectrum. As a reference, blue dotted contours show the results 
combining both $P_0$ and 
$P_2$, but (incorrectly) neglecting the covariance between monopole and 
quadrupole spectra, i.e., 
$\widetilde{\rm Cov}^{02}=\widetilde{\rm Cov}^{20}=0$. 
\label{fig:2D_error}}
\end{figure}

As a pedagogical example, let us first examine how 
the lower-multipole spectra can constrain 
the parameters $D_A$, $H$, and $f$.
Fig.~\ref{fig:2D_error} shows the two-dimensional contour of 
the $1$-$\sigma$ (68\% C.L.) errors on $(D_A,H)$ (bottom-left), $(D_A,f)$ 
(top-left), and $(f, H)$-planes (bottom right). Here, 
the Fisher matrix is computed adopting the model of redshift-space power 
spectrum (\ref{eq:new_pk_gal}) up to $k_{\rm max}=0.2h$Mpc$^{-1}$. 

The magenta solid and cyan dashed lines respectively represent the 
constraints coming from the monopole ($P_0$) and 
quadrupole ($P_2$) power spectrum alone. 
As anticipated, only the single multipole spectrum cannot provide 
useful information to simultaneously constrain $D_A$, $H$, and $f$. 
In particular, for the constraints on $D_A$ and $H$,  
there appear strong degeneracies, and the error ellipses are 
much elongated and inclined.  These behaviors are basically deduced from 
the Alcock \& Paczynski effect, and are consistent with the facts that 
the monopole spectrum is rather sensitive to the combination $(D_A^2/H)$, 
while the quadrupole spectrum is sensitive to $(D_A\,H)$ 
(e.g., \cite{Padmanabhan:2008ag}). On the other hand, combining monopole 
and quadrupole greatly improves the constraints (indicated by blue, 
outer shaded region) not only on $D_A$  and $H$, but also on growth-rate 
parameter $f$. This is because the degeneracies between 
the parameters $D_A$ and $H$ 
constrained by the monopole differ from that by the quadrupole,  
and thus the combination of these two spectra 
leads to a substantial reduction of the size of error ellipses. Further, 
the growth-rate parameter is proportional to the strength of redshift 
distortion, and can be determined by the 
quadrupole-to-monopole ratio. Although the measurement of 
the galaxy power spectrum alone merely gives a constraint on $\beta=f/b$,  
provided the accurate CMB measurement for power spectrum normalization, 
we can separately determine the growth-rate parameter. Note that the 
combination of monopole and hexadecapole spectra also provides a way to 
determine the growth-rate parameter (red shaded region),  
although the error on $f$ is a bit larger due to the small amplitude of 
hexadecapole spectrum. 

For comparison, Fig.~\ref{fig:2D_error} shows the forecast constraints 
obtained from the full 2D power spectrum (green, inner shaded region). 
Further, we plot the results combining the monopole and quadrupole spectra, 
but neglecting the covariance between $\ell=0$ and $\ell=2$, i.e., 
$\widetilde{\rm Cov}^{02}=\widetilde{\rm Cov}^{20}=0$ (blue, dotted 
lines). Clearly, using a full 2D shape of the redshift-space power 
spectrum leads to a tighter constraint, and the area of the two-dimensional 
error is reduced by a factor of $1.6-18$, compared with the constraints 
from the monopole and quadrupole spectra. The results indicate that 
the contribution of the higher multipoles is very important,  
and the additional information from quadrupole and hexadecapole spectra, 
each of which puts a different parameter degeneracy, 
seems to play a dominant role in improving the constraints. 
On the other hand, for joint constraints from the monopole and quadrupole,
a role of the covariance $\widetilde{\rm Cov}^{02}$ or 
$\widetilde{\rm Cov}^{20}$ seems less important, and one may naively treat 
monopole and quadrupole power spectra as statistically independent 
quantities. However, these results are partially due to the properties of the 
galaxy samples characterized by several parameters, and may be altered in 
different assumptions or survey setup. 
This point will be investigated in some details in next subsection.

\subsection{Figure-of-Merit}
\label{subsec:FoM}

\begin{figure*}

\vspace*{-1.5cm}

\begin{center}
\includegraphics[width=10cm,angle=0]{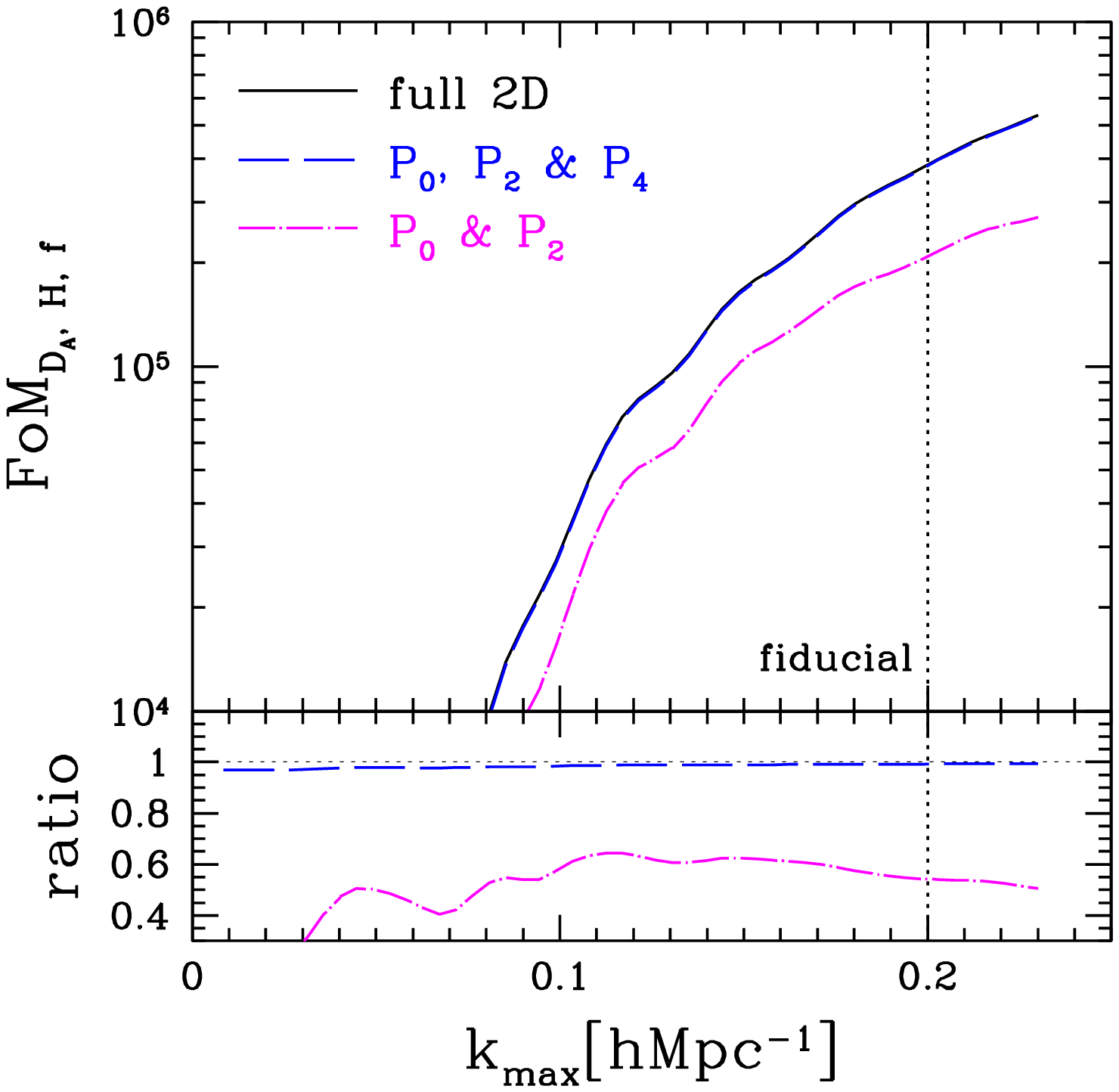}
\hspace*{-2.4cm}
\includegraphics[width=10cm,angle=0]{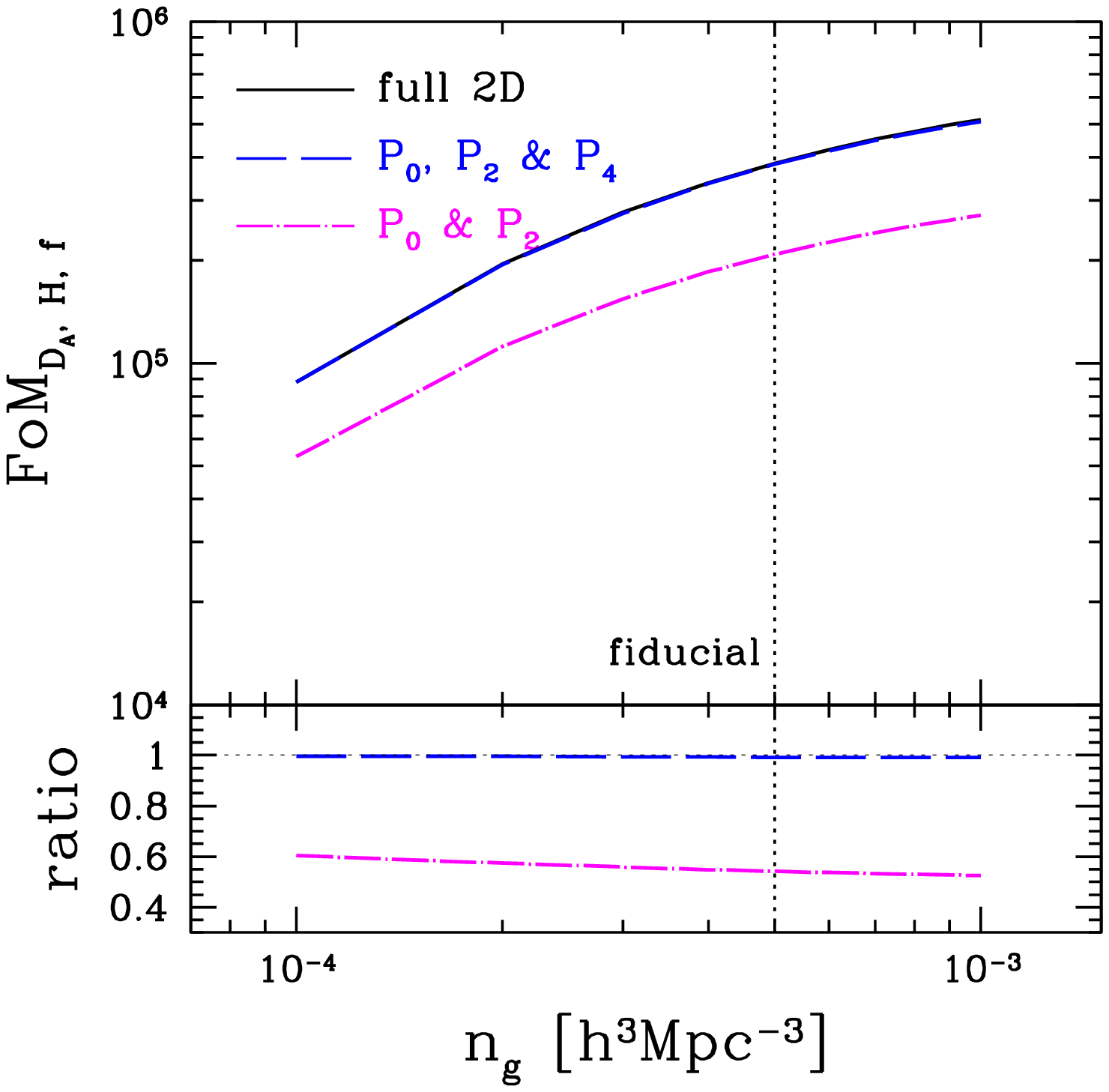}

\vspace*{-2.4cm}

\includegraphics[width=10cm,angle=0]{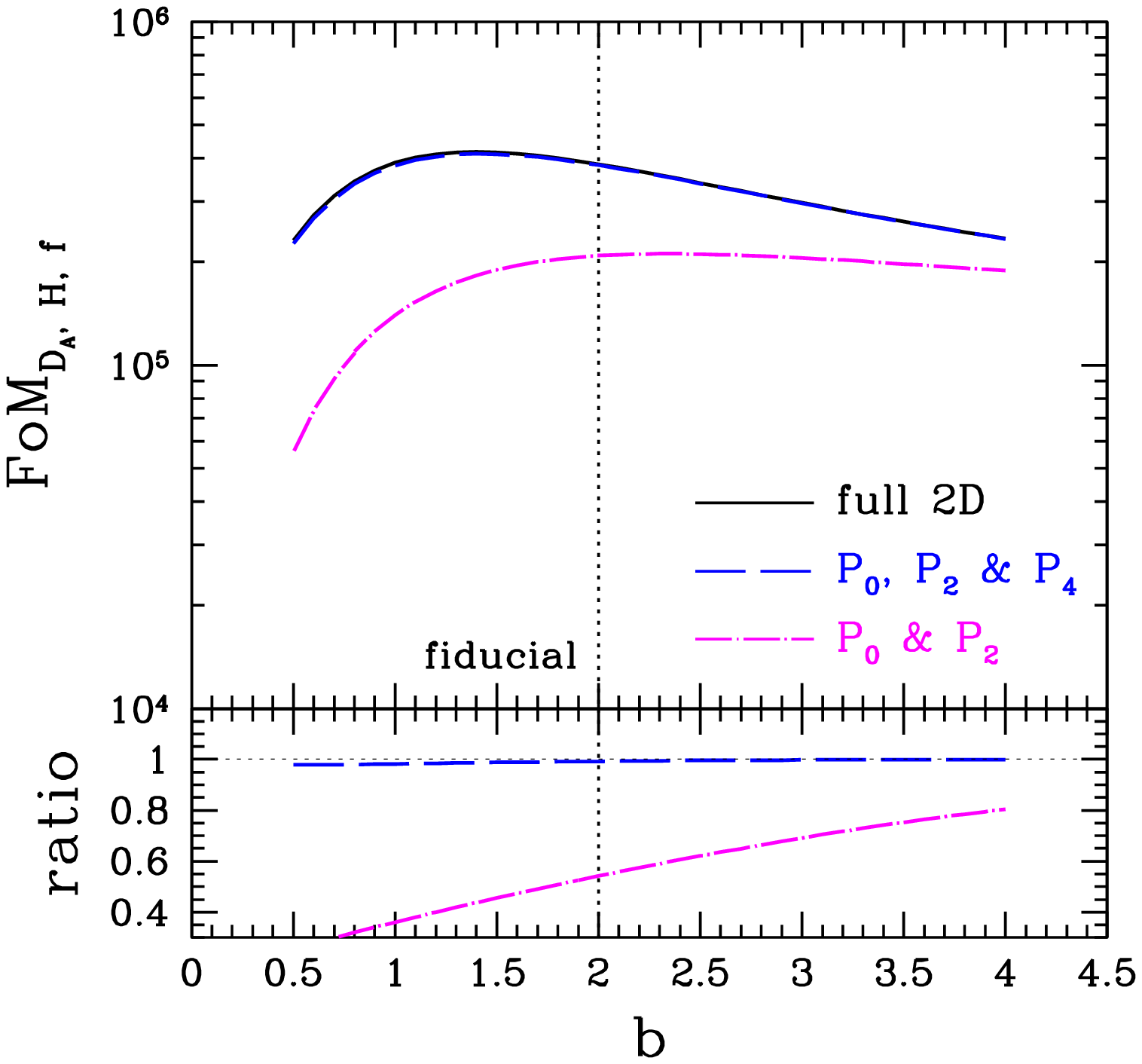}
\hspace*{-2.4cm}
\includegraphics[width=10cm,angle=0]{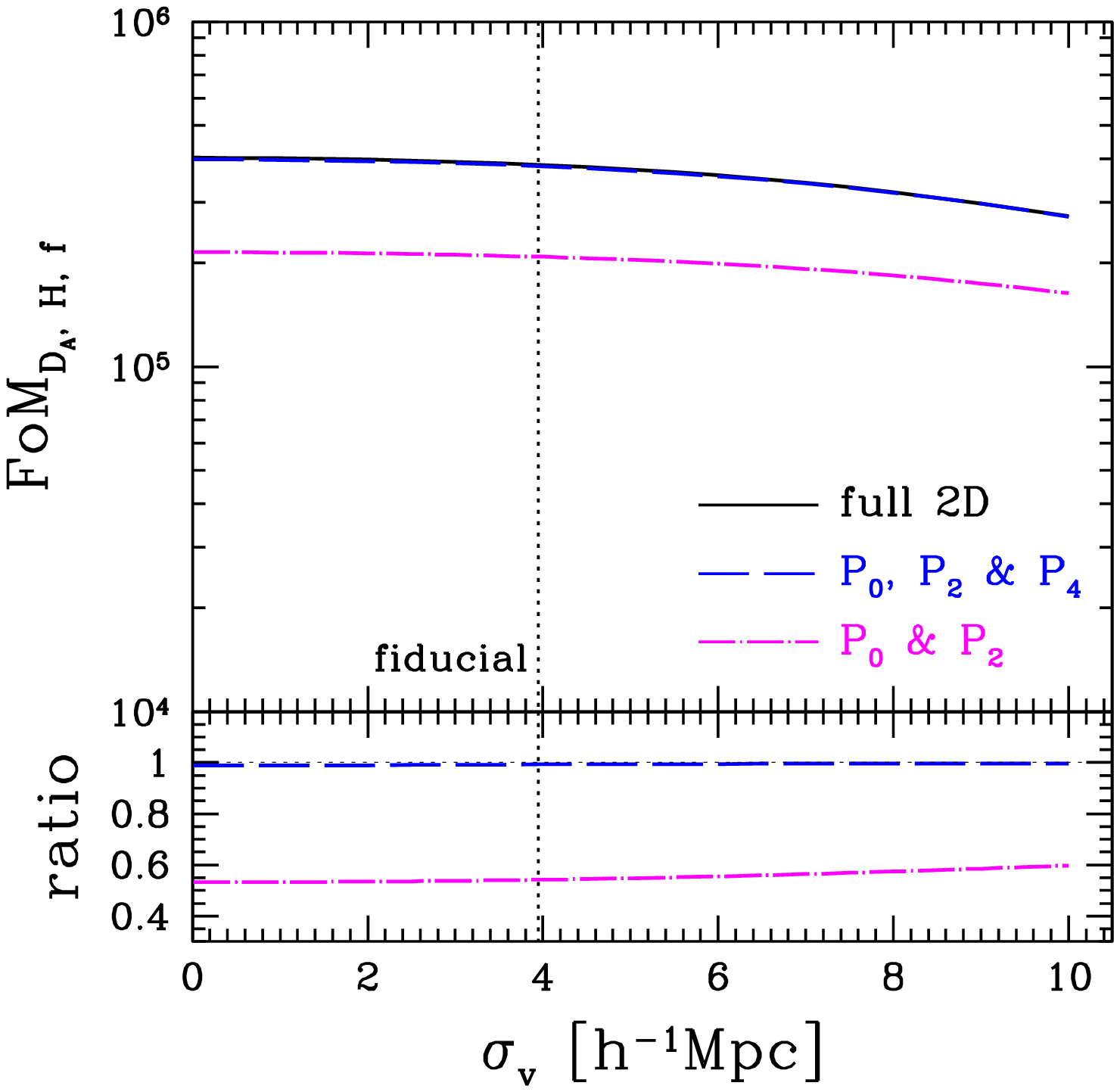}
\end{center}

\vspace*{-1.2cm}

\caption{Figure-of-merit (FoM) on the parameters 
$D_A$, $H$, and $f$ defined by Eq.~(\ref{eq:def_FoM}), 
as functions of $k_{\rm max}$ (top-left), $\overline{n}_{\rm gal}$ (top-right), 
$b$ (bottom left), and $\sigmav$ (bottom right), assuming a hypothetical 
galaxy survey at $z=1$ with volume $V_s=4\,h^{-3}$Gpc$^3$. 
In each panel, solid lines are the results 
obtained from the full 2D 
power spectrum, while 
the dashed and dot-dashed lines represent the FoM from the combination of 
the multipole spectra (dot-dashed: $P_0$ \& $P_2$, dashed: 
$P_0$, $P_0$, \& $P_4$). The bottom panels 
show the ratio of FoM normalized by the one obtained 
from the full 2D spectrum. 
Note that except the parameter along the 
horizontal axis, the fiducial values of the model parameters are 
set to $k_{\rm max}=0.2\,h$Mpc$^{-1}$, 
$n_{\rm g}=5\times10^{-4}\,h^3$Mpc$^{-3}$, $b=2$, and 
$\sigmav=3.95\,h^{-1}$Mpc, indicated by the vertical dotted lines. 
\label{fig:FoM}}
\end{figure*}

\begin{figure}[t]

\vspace*{-2.5cm}

\begin{center}
\includegraphics[width=10cm,angle=0]{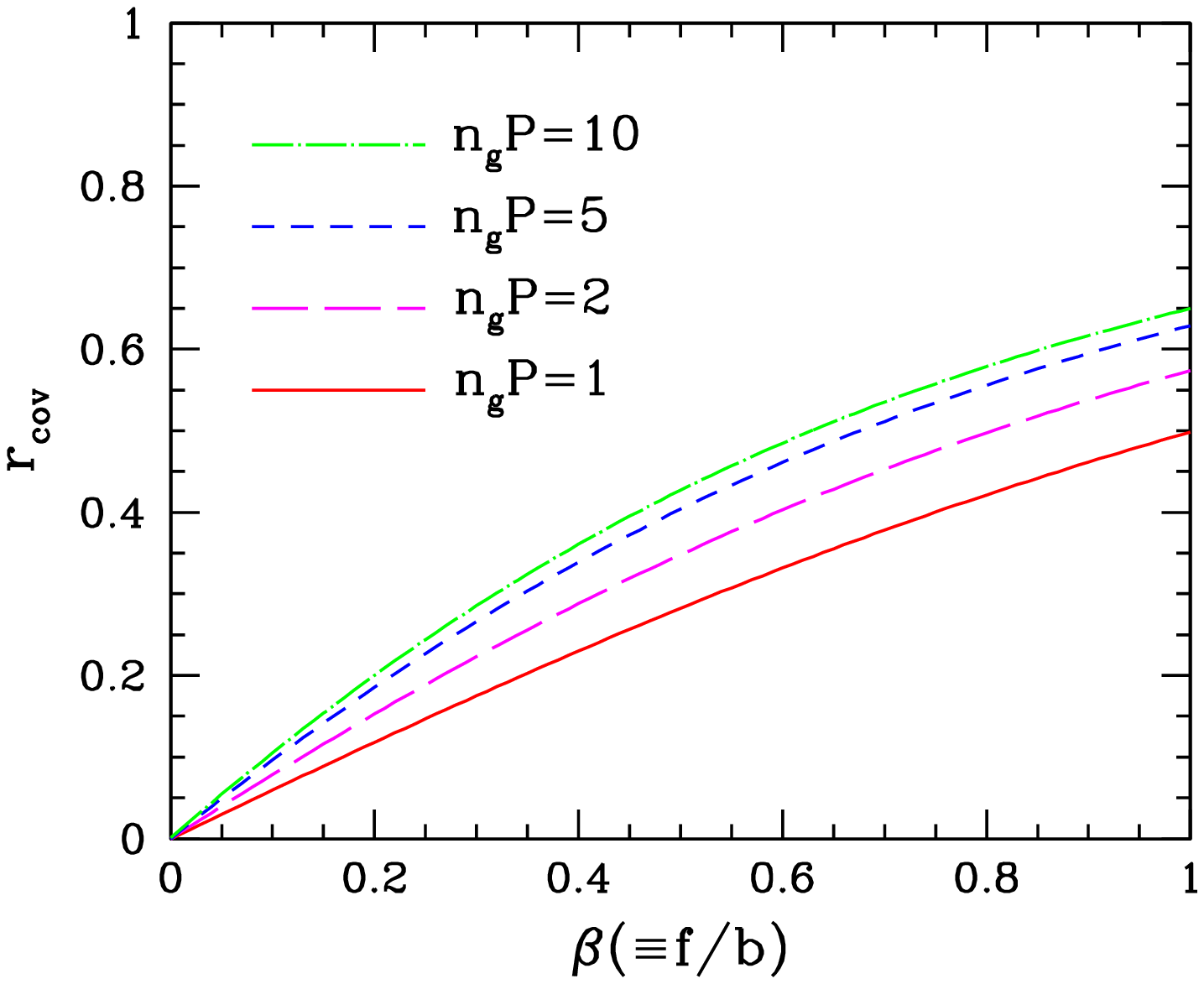}
\end{center}

\vspace*{-2.0cm}

\caption{Correlation coefficient for the covariance, 
$r_{\rm cov}=\widetilde{\rm Cov}^{0,2}/[\widetilde{\rm Cov}^{0,0}
\widetilde{\rm Cov}^{2,2}]^{1/2}$, 
as function of $\beta\equiv f/b$. The plotted results are obtained 
based on the linear theory, in which the coefficient $r_{\rm cov}$ 
depends on the power spectrum amplitude relative to the 
shot-noise contribution, $n_{\rm g}\,P$, as well as $\beta$. 
The solid, long-dashed, short-dashed, and dot-dashed lines respectively 
indicate the results with $n_{\rm g}\,P=1$, $2$, $5$, and $10$.
\label{fig:rcross}}
\end{figure}

We here study the dependence of galaxy samples or survey setup on 
the forecast results for parameter constraints. To do this, 
it is useful to define the Figure-of-Merit (FoM):
\begin{align}
\mbox{FoM}\equiv 
\frac{1}{\sqrt{\mbox{det}\widetilde{\mbox{\boldmath$F$}}^{-1}}}, 
\label{eq:def_FoM}
\end{align}
where the matrix $\widetilde{\mbox{\boldmath$F$}}^{-1}$ is the 
$3\times3$ sub-matrix, whose elements are taken from the  
inverse Fisher matrix $\mbox{\boldmath$F$}^{-1}$ associated with the 
parameters $D_A$, $H$, and $f$. The FoM quantifies the improvement 
of the parameter constraints, and inversely proportional to 
the product of one-dimensional marginalized errors, i.e., 
$\mbox{FoM}\propto 1/\{\sigma(D_A)\sigma(H)\sigma(f)\}$.

Fig.~\ref{fig:FoM} shows the dependence of FoM on the 
properties of the galaxy samples characterized by the number density 
$n_{\rm g}$ (top-right), bias parameter $b$ (bottom-left), and 
one-dimensional velocity dispersion $\sigmav$ (bottom-right). 
Also, in top-left panel, we show the FoM as a function of maximum 
wavenumber $k_{\rm max}$ used in the parameter estimation study. Note that 
in plotting the results, the other parameters are kept fixed to the 
canonical values. 
The upper part of each panel plots the three different lines, and 
shows how the FoM changes depending on the choice or combination 
of power spectra used in the analysis: combining monopole ($P_0$) 
and quadrupole ($P_2$) spectra (magenta, dot-dashed); combining 
three multipole spectra, $P_0$, $P_2$ and $P_4$ (blue, long-dashed); 
using full 2D spectrum $P(k,\mu)$ (black, solid). On the other hand, 
the lower part of each panel plot the ratio of FoM normalized by 
the one for the full 2D spectrum.

In principle, using the full 2D spectrum gives the tightest constraints 
on $D_A$, $H$, and $f$, but an interesting 
point here is that almost equivalent FoM to the one for the full 2D spectrum 
is obtained even from a partial information with 
the lower-multipole spectra $P_0$, $P_2$ and $P_4$. This is 
irrespective of the choice of the parameters for galaxy samples.  
Although the result may rely on the model of redshift 
distortion adopted in this paper, recalling the fact that the non-vanishing 
multipole spectra higher than $\ell\gtrsim6$ arise only from 
the non-linear effects through the gravitational evolution and 
redshift distortion, the cosmological model dependence encoded in 
these higher multipoles is expected to be very weak, partly due to the 
low signal-to-noise ratio. In this sense, 
the result in Fig.~\ref{fig:FoM} seems reasonable.

Now, turn to focus on the FoM from the combination of $P_0$ and $P_2$. 
Fig.~\ref{fig:FoM} indicates that except for the case varying the 
bias $b$, the resultant FoM 
shows a monotonic dependence on the parameters. 
As a result, the ratio of FoM shown in the lower part of the panels 
is nearly constant around $0.4-0.6$.  
As for the variation of bias parameter, 
the non-monotonic dependence of the FoM is 
basically explained by the two competitive effects. 
That is, as increasing $b$, 
while the power spectrum amplitude increases and signal-to-noise ratio 
is enhanced, the clustering anisotropies due to the redshift distortion 
controlled by the quantity $\beta$ are gradually reduced. 
Hence, for some values of 
$b$, FoM becomes maximum. A noticeable point is that 
the ratio of FoM for the monopole and quadrupole becomes gradually 
increased as the clustering bias becomes large. At $b\sim4$, 
the ratio of FoM reaches at $0.8$, indicating 
most of the cosmological information contained in the 
hexadecapole and higher multipoles is lost, 
and signals coming from the monopole and quadrupole spectra becomes
dominated.

The reason for this behavior is presumably due to the covariance 
between the multipole spectra, $\widetilde{\mbox{Cov}}^{\ell\ell'}$. 
In linear regime, the covariance neglecting the shot noise contribution
is determined by the galaxy power spectrum in real space 
and parameter $\beta=f/b$, and the off-diagonal component 
$\widetilde{\mbox{Cov}}^{02}=\widetilde{\mbox{Cov}}^{20}$ is roughly 
proportional to $\beta$. Thus, as increasing the clustering bias $b$ while 
keeping the growth-rate parameter, the covariance 
$\widetilde{\mbox{Cov}}^{02}$ becomes smaller, and the 
monopole and quadrupole power spectra become statistically independent. 
To see this more explicitly, we define 
\begin{align}
r_{\rm cov}= \frac{\widetilde{\rm Cov}^{0,2}}
{[\widetilde{\rm Cov}^{0,0}\widetilde{\rm Cov}^{2,2}]^{1/2}}. 
\label{eq:r_cov}
\end{align}
In Fig.~\ref{fig:rcross}, taking account of the shot noise contribution, 
the quantity $r_{\rm cov}$ is plotted against the parameter $\beta$. 
Here, we used the linear theory to calculate $\widetilde{\rm Cov}^{\ell\ell'}$. 
Fig.~\ref{fig:rcross} implies that in our fiducial setup with $f=0.858$, 
$r_{\rm cov}$ becomes $\lesssim0.2$ for the bias $b=4$. Since the 
smaller values of $\beta$ also suppress the Kaiser effect in the 
covariances $\widetilde{\rm Cov}^{00}$ and $\widetilde{\rm Cov}^{22}$, 
the constraints from the monopole and quadrupole spectra is 
relatively improved.

The result suggests that even the partial information with monopole and 
quadrupole spectra 
still provides a fruitful constraint on $D_A$, $H$ and $f$, depending on 
the survey setup. In this respect, a benefit to use these power spectra 
should be further explored. As a next step, we will discuss the robustness 
of the parameter constraints against the systematic biases.

\subsection{Impact of systematic biases}
\label{subsec:FoB}

\begin{figure*}[t]

\vspace*{-0.5cm}

\begin{center}
\includegraphics[width=8.9cm,angle=0]{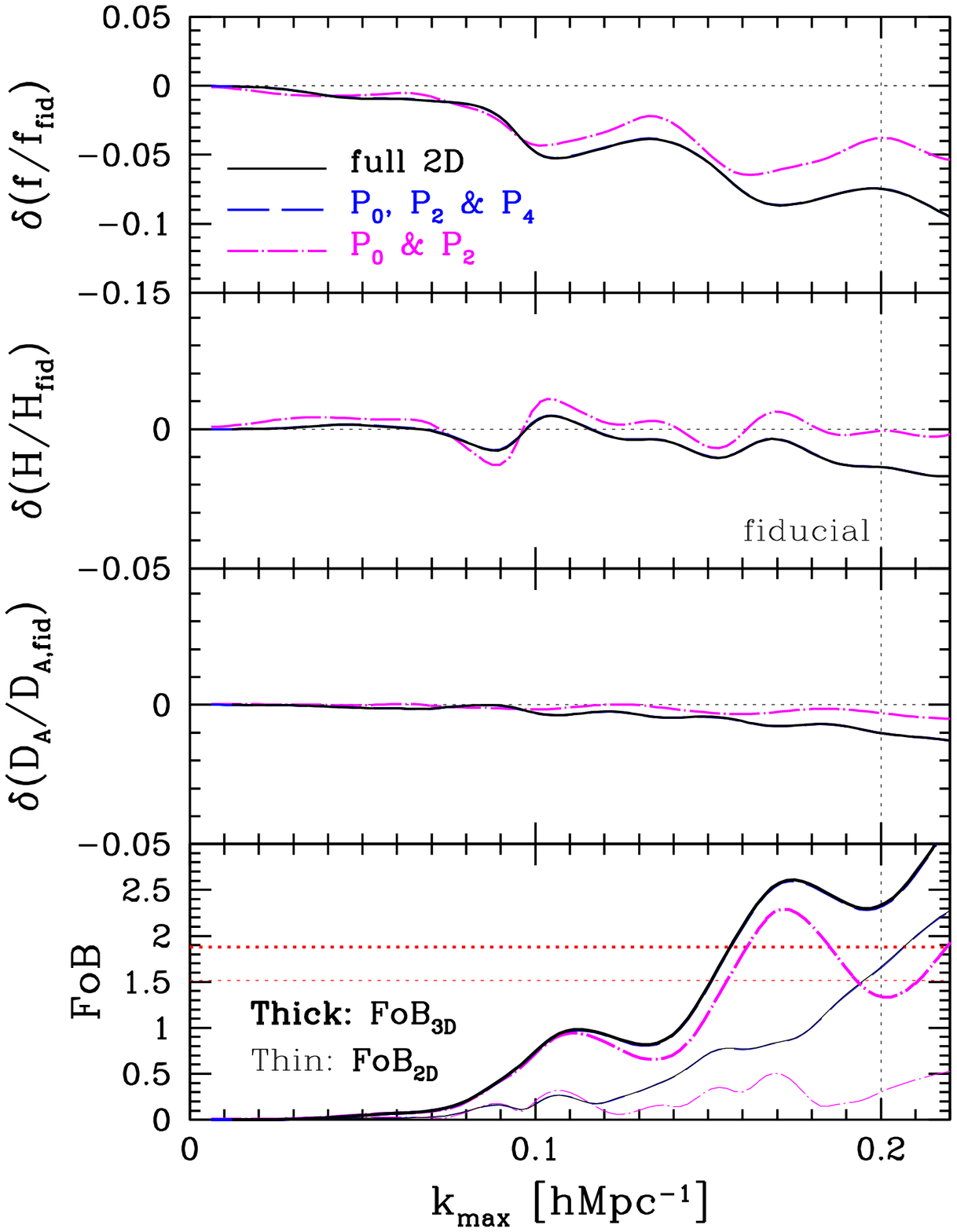}
\includegraphics[width=8.9cm,angle=0]{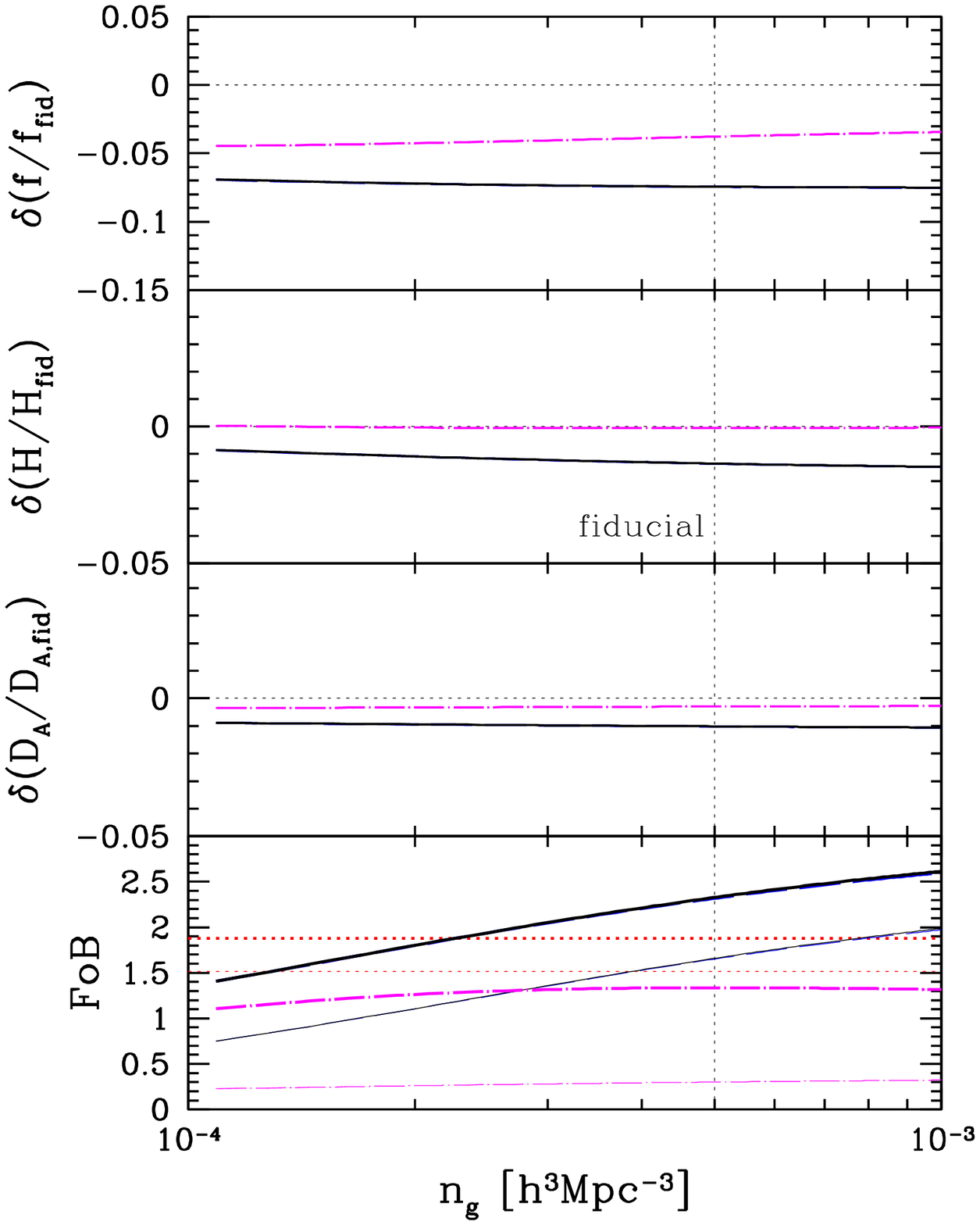}

\vspace*{0.2cm}

\includegraphics[width=8.9cm,angle=0]{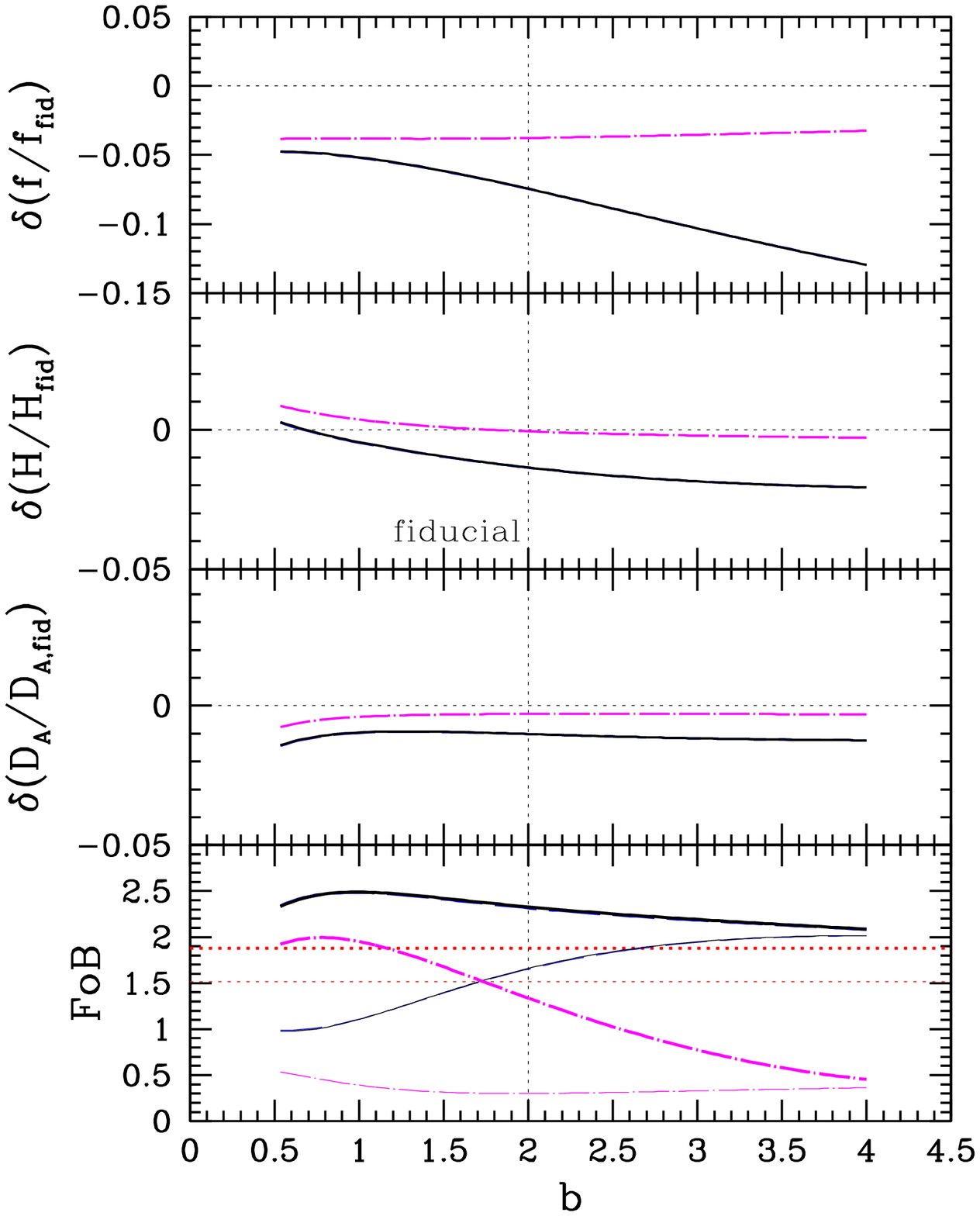}
\includegraphics[width=8.9cm,angle=0]{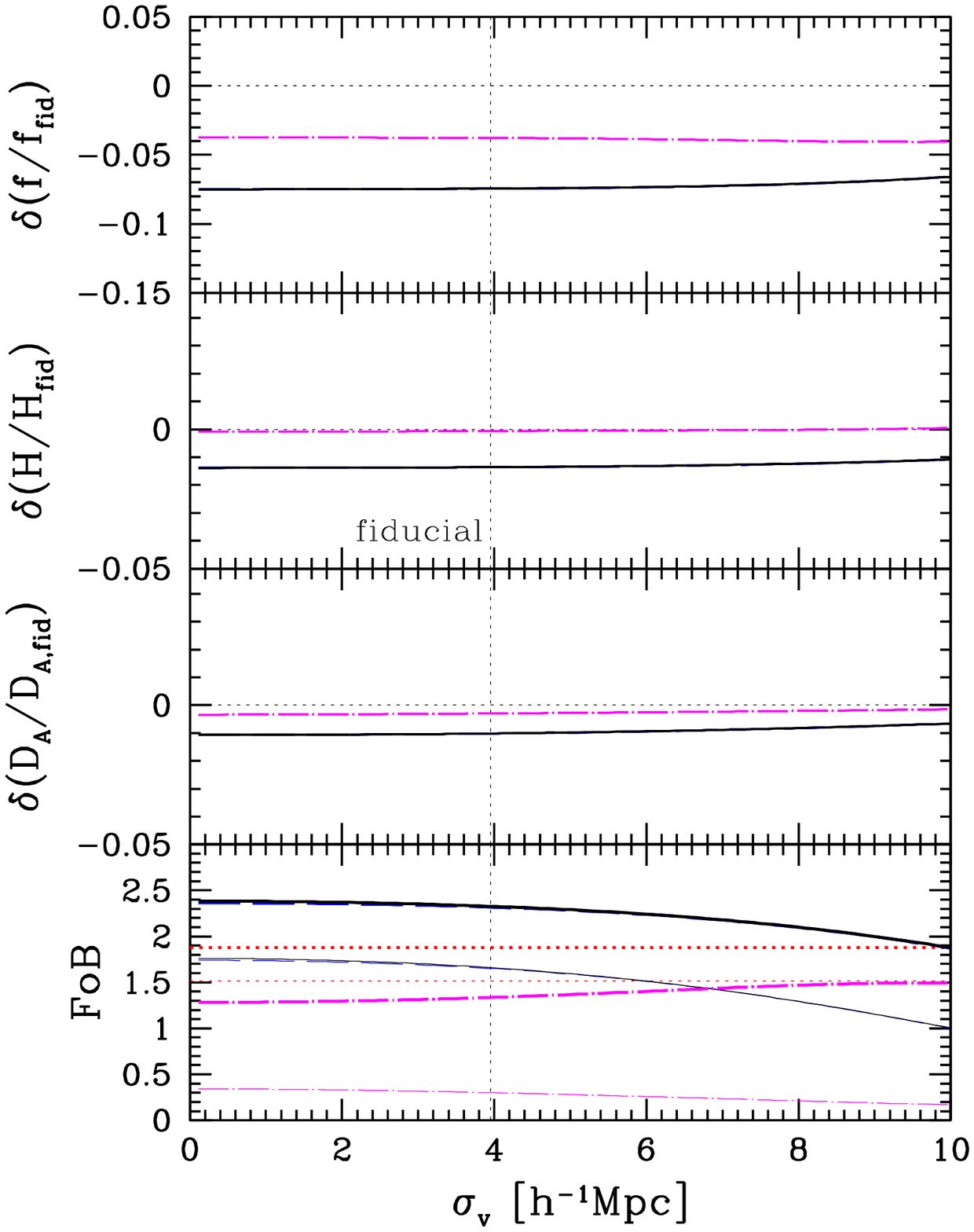}
\end{center}

\vspace*{-0.5cm}

\caption{Systematic biases for best-fit values of parameters 
$f$, $D_A$ and $H$ and Figure-of-Bias as function of $k_{\rm max}$ (top-left), 
$n_{\rm g}$ (top-right), $b$ (bottom-left), and $\sigmav$ (bottom-right).  
These are the estimates adopting the 'incorrect' model of redshift-space 
power spectrum, in which we ignore the small correction terms, $A$ and $B$.  
In bottom panels of each figure, thick and thin lines respectively show 
the FoB in three and two-dimensions, i.e., $(D_A,H,f)$ and $(D_A,H)$. 
The dotted lines indicates the $1\mbox{-}\sigma$ significance of the 
deviation relative to the statistical error. Note that 
the shift of the best-fit parameters remains unchanged irrespective of the 
survey volume $V_s$, while the FoB given here represents the specific results 
with the survey volume $V_s=4h^{-3}$Gpc$^3$. The fiducial values of the 
model parameters used in the calculation are the same as 
in Fig.~\ref{fig:FoM} (indicated by vertical dotted lines), 
except the parameter along the horizontal axis. 
\label{fig:FoB}}
\end{figure*}

Among various envisaged systematics that affect the 
parameter constraints, the incorrect assumption for the theoretical 
template of power spectra may seriously lead to a bias in the best-fit 
parameters. There are several routes to produce an incorrect 
theoretical template; incorrect model of redshift distortion and/or 
non-linear gravitational evolution, wrong prior information for 
cosmological parameters, and improper parametrization for 
galaxy bias. 
In this subsection, we specifically examine the first and second cases.  
We first discuss the incorrect model of redshift distortion, and 
quantify the size of the systematic bias in the best-fit parameter. 
The influence of the wrong prior information will be discussed in next 
subsection.

\subsubsection{Systematic biases from a wrong model of redshift distortion}
\label{subsubsec:FoB}

Let us first discuss the impact of incorrect model of redshift distortion 
on the parameter estimation. To be precise, we 
consider the small discrepancy in the theoretical 
template for redshift-space power spectrum (\ref{eq:new_pk_gal}),  
and estimate the systematic biases from Eq.(\ref{eq:systematic_bias}).
Fig.~\ref{fig:FoB} shows 
the systematic biases caused by the incorrect model template neglecting 
the $A$ and $B$ terms. 
We plot the results by varying the model parameters, $k_{\rm max}$ (top-left), 
$n_{\rm g}$ (top-right), 
$b$ (bottom-left), and $\sigmav$ (bottom-right), around the fiducial 
values. In each panel, the first three panels from the top plot the 
deviation of the best-fit value from the fiducial one, $\delta f$, 
$\delta D_A$, and $\delta H$, normalized by their fiducial values. 
On the other hand, the lowest panel shows the Figure-of-Bias (FoB),  
which represents the statistical significance of systematic biases  
relative to the statistical errors, 
defined by \cite{2009PhRvD..80b3003J,2009ApJ...696..775S}:
 \begin{align}
 \mbox{FoB}\equiv 
 \left(\sum_{i,j}\delta \theta_i\widetilde{\mbox{\boldmath$F$}'}_{ij}\delta 
\theta_j\right)^{1/2}
\label{eq:FoB}
 \end{align}
Note that the matrix $\widetilde{\mbox{\boldmath$F$}'}_{ij}$ is the 
same inverse of the sub-matrix $\widetilde{\mbox{\boldmath$F$}}_{ij}^{-1}$
as defined in Eq.~(\ref{eq:def_FoM}), but with the Fisher matrix 
obtained from the incorrect template. With the above definition,  
the FoB squared simply reflects the $\Delta\chi^2$ for 
the true values of the parameters relative to the 
biased estimate of the best-fit values \cite{2009ApJ...696..775S}. 
Thus, in the cases with three parameters, if the FoB exceeds $1.88$ 
(indicated by the red, thick dotted lines), the true values of the 
parameters would go outside the 1-$\sigma$ ($68$\%C.L.) error ellipsoid 
of the biased confidence region. Notice that 
the shift of best-fit parameters remains unchanged irrespective of 
the survey volume $V_s$, while the FoB is proportional to $V_s^{1/2}$.

Fig.~\ref{fig:FoB} shows that the biases in the distance information, 
$\delta D_A$ and $\delta H$, are basically small and reach 
$1-2$\% at most, but the bias in the growth-rate parameter, $\delta f$, 
is rather large. Hence, the behaviors of the FoBs indicated by the 
thick lines are mostly dominated by the error and bias in the 
growth-rate parameter. As a result, for some ranges of parameters, 
the expected FoB using a full-shape information (black solid, labeled 
as 'full 2D') tends to exceed the critical value, $1.88$. This is 
true even if we marginalize over $f$ and just focus on the 
distance information $D_A$ and $H$, depicted as thin lines 
in the lowest panels (labeled as 'FoB$_{\rm 2D}$'). Note that in the 
case of two parameters, the true values of $D_A$ and $H$ are ruled out 
at $1$-$\sigma$ level if FoB exceeds $1.52$ (red, thin dotted lines).

On the other hand, if we use the information obtained only from 
the monopole and quadrupole spectra (magenta, dot-dashed lines), 
the systematic biases are significantly reduced, and 
the resultant FoBs are well within the critical values except for 
unrealistic case with a large $\sigmav$ or anti-bias $b\lesssim1$. If 
we are just interested in $D_A$ and $H$ marginalized over $f$, the FoB
becomes substantially smaller, and would be far below the 
critical value $1.52$, even for a large galaxy survey with
$V_s\lesssim40h^{-3}$Mpc$^3$. Therefore even the partial information from 
the monopole and quadrupole spectra is helpful 
and rather robust against the systematic biases than the full 2D information. 
Although the figure-of-merit for the constraints on $D_A$, $H$ and $f$ 
would be degraded, the reduction of FoM is at most factor of $\sim0.6$, 
which can be improved to $\sim0.8$ for highly biased objects 
(see Fig.~\ref{fig:FoM}).

\begin{figure*}[t]
\begin{center}
\includegraphics[width=10.5cm,angle=0]{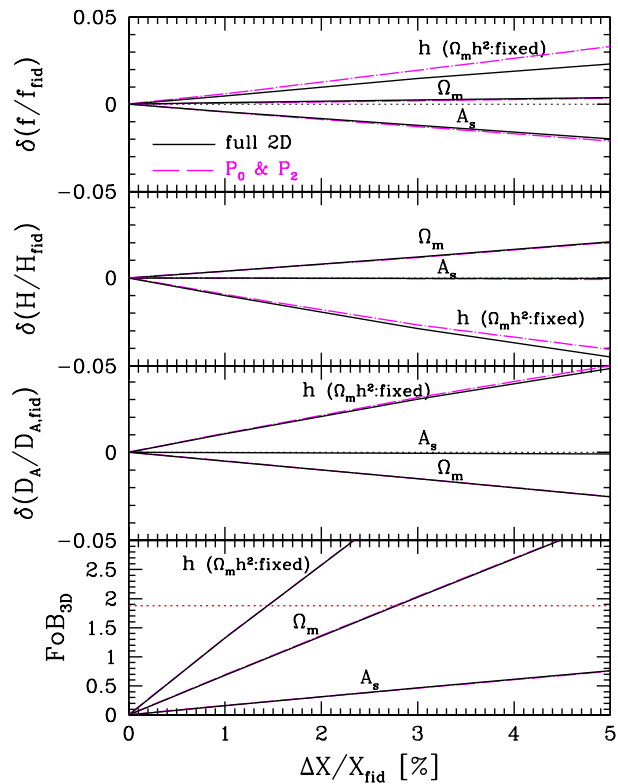}
\end{center}

\vspace*{-0.3cm}

\caption{Systematic biases for the best-fit values of the parameters 
$f$, $D_A$ and $H$, and FoB for these three parameters 
(from top to bottom), adopting the incorrect prior information 
for cosmological parameters in computing 
the template power spectrum; $X=A_s$, 
$\Omega_{\rm m}$, and $h$ ($\Omega_{\rm m}h^2$: fixed).     
The results are plotted against the fractional difference 
between the correct and incorrect values of each cosmological 
parameters, $\Delta X/X_{\rm fid}$. Solid and dashed lines represent 
the results from a full 2D power spectrum and partial information 
with monopole and quadrupole spectra, respectively. Note that in bottom 
panel, the horizontal dotted 
lines indicates the $1\mbox{-}\sigma$ significance of the 
deviation relative to the statistical error.   
\label{fig:FoB_cosmoparams}}
\end{figure*}

Finally, there are several interesting points to be noted. 
One is the oscillatory 
behavior of the systematic biases and FoB shown in the top-left panel.  
This is originated from the acoustic structure of the power spectrum, 
and the result suggests that the bias in the growth-rate parameter 
$\delta f$ is sensitively affected by the BAO measurement. 
Another noticeable feature is a suppression of the FoB 
in the case of three parameters using the monopole and quadrupole spectra,  
which appears at a larger value of the galaxy bias $b$ 
(thick, dot-dashed line in bottom-left panel). This is presumably due to 
the multiple effects that as increasing the clustering bias, the 
systematic bias for the growth-rate parameter tends to be slightly 
reduced, while the constraint on the growth-rate parameter becomes 
gradually weaker. There also appears a similar trend in the case 
using a full 2D spectrum, but the suppression is rather small and FoB never 
falls below the critical value, $1.88$. This is because the 
biased estimate of growth-rate parameter, $\delta f$, 
significantly deviates from the fiducial value, 
as opposed to the case using monopole and quadrupole spectra.

\subsubsection{Systematic biases from incorrect prior information}
\label{subsec:bias_prior}

So far, we have assumed that the underlying cosmological parameters 
necessary to compute the redshift-space power spectrum are whole 
known a priori from the CMB observations such as PLANCK. However, even the 
precision CMB measurement produces some uncertainties in the 
cosmological parameters due to the parameter degeneracy.  
This may give an important source for the incorrect theoretical template for 
redshift-space power spectrum, and leads to a biased estimate of 
$D_A$, $H$, and $f$. 

Fig.~\ref{fig:FoB_cosmoparams} quantifies the size of systematic biases and 
FoB arising from the incorrect assumptions for cosmological parameters. 
Here, we especially focus on the parameters $A_s$, 
$\Omega_{\rm m}$, and $h$ fixing $\Omega_{\rm m}h^2$ constant, and 
plot the sensitivity of the systematic biases to the variation of 
those parameters. Note that in computing the power spectrum, 
we strictly assume the flat cosmological model and the model of 
redshift distortion (\ref{eq:new_pk_gal}) as a fiducial model of 
power spectrum template.

Compared to the results in Sec.~\ref{subsubsec:FoB}, the systematic 
bias in the growth-rate parameter is relatively small, and the 
significance of the biases in the acoustic-scale information 
conversely increases. That is, the best-fit values of the 
parameters $D_A$ and $H$ is rather sensitive to the precision of the 
prior information in the power spectrum template. A noticeable point is 
that this is true irrespective of the choice of the template power spectra
used in the parameter estimation (i.e., full 2D spectrum or 
combination of $P_0$ and $P_2$).  As a result, a percent-level 
precision is generally required for the prior information of cosmological 
parameters, except for the scalar spectral amplitude, 
$A_s$. Through the 
non-linear clustering and/or redshift distortion, a small change in $A_s$ 
alters the power spectrum shape, and it can potentially affect 
the acoustic scale and the clustering anisotropies. However, 
at $z=1$, the non-linear effects on the scales of our interest, 
$k\lesssim0.2h$Mpc$^{-1}$, is rather mild, and the resultant 
impact on the acoustic-scale measurement is extremely small. Hence, 
for a typical survey volume of stage III-class survey with 
$V_s\sim4h^{-3}$Gpc$^3$, 
no appreciable systematic bias might be produced 
from the incorrect prior assumption on $A_s$.

\section{Summary}
\label{sec:summary}

In this paper, we have studied the cosmological constraints 
from the anisotropic BAOs based on the multipole expansion of 
redshift-space power spectrum.  We have derived 
the several formulae for the Fisher analysis using the 
multipole power spectra; Eqs.~(\ref{eq:formula_F_ij}) and 
(\ref{eq:cov_formula}) for the Fisher matrix, and 
Eqs.~(\ref{eq:systematic_bias}) and (\ref{eq:vector_s}) for 
the estimation of systematic biases. We then consider the 
hypothetical galaxy survey of $V_s=4h^{-3}$Gpc$^3$ and $z=1$, 
and discuss the potential power of 
the lower multipole spectra on the cosmological constraints, particularly 
focusing on the parameters $D_A$, $H$ and $f$.

Compared to the analysis with full 2D power spectrum, 
a partial information from the monopole and quadrupole 
power spectra generally degrades the constraints on 
$D_A$, $H$, and $f$. Typically, the constraint is degraded 
by a factor of $\sim1.3$ for each parameter. 
The interesting finding is that adding the information from hexadecapole 
spectra ($P_4$) to that from the monopole and quadrupole spectra 
greatly improves the constraints, and the resultant 
constraints would become almost comparable to those 
expected from the full 2D power spectrum (see Fig.~\ref{fig:FoM}). 
Note also that the situation would be relatively improved depending 
on the properties of galaxy samples, and for highly biased galaxy samples 
with $b\sim4$, the total power of the constraints defined by the 
Figure-of-Merit [FoM, Eq.~(\ref{eq:def_FoM})] 
can reach $\sim80\%$ of the one expected from the full 2D power spectrum.

We have also investigated the impacts of systematic biases on the 
best-fit values of $D_A$, $H$ and $f$. 
The incorrect model of redshift distortion tends to produce a large
systematic bias in the growth-rate parameter, and the size of biases 
would be rather significant for the analysis with full 2D spectrum. 
An interesting suggestion is that the situation would be greatly relaxed 
if we only use the combination of monopole and quadrupole spectra, and the 
estimated value of Figure-of-Bias defined by Eq.~(\ref{eq:FoB}) is mostly 
below the critical value for stage-III class surveys (Fig.~\ref{fig:FoB}). 
In this respect, the analysis with partial information from monopole and 
quadrupole may be still helpful in cross-checking the results derived 
from the full 2D power spectrum. On the other hand, 
wrong prior assumption of cosmological parameters in computing 
the template power spectrum severely affects the acoustic-scale 
determination, and a percent-level precision is required for 
the prior information in order to avoid a large systematic biases
on $D_A$ and $H$ (Fig.~\ref{fig:FoB_cosmoparams}). 
This is true irrespective of the choice of template power spectra 
used in the analysis.

Finally, we note that the assumptions and situations considered in the paper 
are somewhat optimistic or too simplistic, and a more careful study is 
needed for a quantitative parameter forecast. One critical aspect is 
the modeling of the galaxy power spectrum. 
In reality, the assumption of linear and deterministic galaxy biasing 
is idealistic, and the scale-dependence or non-linearity/stochasticity  
of the galaxy biasing should be consistently incorporated into the 
theoretical template of redshift-space power spectrum. Although 
this is tiny effect for the scale of our interest, 
the distance information, $D_A$ and $H$, is 
rather sensitive to a slight modification of the acoustic structure in 
the power spectrum, and results in this paper 
might be somehow changed. A more elaborate modeling for power spectrum 
is thus quite essential.

\begin{acknowledgements}
AT is supported by a Grant-in-Aid for Scientific 
Research from the Japan Society for the Promotion of Science (JSPS) 
(No.~21740168). TN is supported by JSPS. 
SS is supported by JSPS through research fellowships and Excellent Young
 Researchers Overseas Visit Program (No.21-00784). 
This work was supported in part by 
Grant-in-Aid for Scientific Research on Priority Areas No.~467 
``Probing the Dark Energy through an Extremely Wide and Deep Survey with 
Subaru Telescope'', JSPS Core-to-Core Program ``International 
Research Network for Dark Energy'', and World Premier International Research 
Center Initiative (WPI Initiative), MEXT, Japan.
\end{acknowledgements}


\end{document}